\documentclass{mn2e}
\usepackage{times}
\input{psfig.sty}
\newif\ifAMStwofonts

\title{Using radio emission to detect isolated and quiescent accreting
black holes}

\author[Maccarone] {Thomas J. Maccarone\\ Astronomical Institute
``Anton Pannekoek'', University of Amsterdam, Kruislaan 403, 1098 SJ,
Amsterdam, The Netherlands}

\date{}

\begin{document}

\maketitle

\label{firstpage}

\def\simlt{\mathrel{\rlap{\lower 3pt\hbox{$\sim$}}
        \raise 2.0pt\hbox{$<$}}}
\def\simgt{\mathrel{\rlap{\lower 3pt\hbox{$\sim$}}
        \raise 2.0pt\hbox{$>$}}}

\input epsf

\begin{abstract}

We discuss the implications of new relations between black holes'
masses, X-ray luminosities and radio luminosities, as well as the
properties of the next generation of radio telescopes, for the goal of
finding isolated accreting black holes.  Because accreting black holes
have radio-to-X-ray flux ratios that increase with decreasing
luminosity in Eddington units, and because deep surveys over large
fields of view should be possible with planned instrumentation such as
LOFAR, radio surveys should be significantly more efficient than X-ray
surveys for finding these objects.  

\end{abstract}

\begin{keywords}
binaries:close -- ISM:jets and outflows -- black hole physics --radio
continuum:stars -- dark matter -- cosmic rays
\end{keywords}

\section{Introduction}

It is generally believed that about $10^8-10^9$ black holes of $\sim
10M_\odot$ exist in the Milky Way (Shapiro \& Teukolsky 1983; van den
Heuvel 1992; Samland 1998).  On the other hand, only $\sim20$ strong
black hole candidates are known.  Most of these are black holes in
X-ray binary systems, where the black holes are first detected through
their X-ray emission, and their masses are estimated through
spectroscopic measurements of their rotational velocities (see
McClintock \& Remillard 2005 and references within) while a smaller
fraction of these systems have been detected as the lenses in
microlensing events (Bennett 2001; Mao et al. 2002).

The population of black hole systems that accretes from a binary
companion is likely to be a small fraction of the total number of
black holes in the Galaxy, but due to the relative sensitivity of
X-ray telescopes compared with microlensing surveys, our understanding
of the Galaxy's black hole distribution is dominated by the black
holes in X-ray binaries.  This has prompted several groups to suggest
means for identifying isolated black holes through other means (see
e.g. Fujita et al. 1998; Armitage \& Natarajan 1999; Agol \&
Kamionkowski 2002 and references within).  Furthermore, the
microlensing surveys conducted to date are not sensitive to
intermediate mass black holes.  In this paper, we apply recent
advances in our understanding of accretion and jet production in low
luminosity systems.  We show that isolated black holes accreting from
the interstellar medium, and faint quiescent X-ray binaries accreting
from donor stars, are likely to be discovered in planned radio
surveys.  We show that if a substantial number of intermediate mass
black holes exist in the Galaxy, they should be even easier to detect
in the radio than stellar mass black holes will be.  Finally, we
consider the possibility that these systems will contribute
significantly to the kinetic power and the cosmic ray luminosity of
the Galaxy.

\section{Observational appearance of faint accreting black holes}

In the few years that have passed since the last attempts to predict
how one might best go about looking for the lowest luminosity black
hole accretors, several advances have been made in our understanding
of the observational appearances of such systems.  Most specifically,
our understanding of the relationship between the radio and X-ray
fluxes of accreting black holes has changed rapidly.  Hannikainen et
al. (1998) first showed that there was likely to be a positive
correlation between X-ray and radio flux in several measurements of
the variable X-ray binary, GX~339-4.  Corbel et al. (2000) verified
this result and finding that $L_R$$\propto$$L_X^{0.7}$, where $L_R$ is
the radio luminosity and $L_X$ is the X-ray luminosity. Gallo, Fender
\& Pooley (2003) showed that this relationship applied universally to
X-ray binaries when in the low/hard spectral state, which is typically
found below about 2\% of the Eddington limit (Maccarone 2003) and
occasionally up to about 10\% of the Eddington ratio due to hysteresis
effects (Maccarone \& Coppi 2003), and which shows an X-ray spectrum
dominated by a hard power law with a cutoff at a few hundred keV (see
McClintock \& Remillard 2005 for a discussion of spectral state
nomenclature).  It was then found that active galactic nuclei (AGN)
also follow this relationship between radio and X-ray luminosities,
with an additional dependence on the mass of the black hole, such that
$L_X\propto$$M_{BH}^{0.8}$ (Merloni, Heinz \& DiMatteo 2003; Falcke,
K\"ording \& Markoff 2004), which is likely to be the result of the
difference in the size of the photospheres of black holes of different
masses (see e.g. Falcke \& Biermann 1996; Heinz \& Sunyaev 2003).  

A secondary consequence of the $L_R\propto$$L_X^{0.7}$ relationship is
that the kinetic power of the jets takes up an increasing fraction of
the total accretion power as the luminosity of a system decreases.
Fender, Gallo \& Jonker (2003) showed that this relationship can be
explained if the kinetic power taken away by the jet goes as
$L_X^{0.5}$, which in turn requires that $L_X$ goes as the square of
the mass accretion rate at low luminosities, because the kinetic power
put into the jet dominates the total accretion power.  From these
relationships, it becomes clear that the fainter and more massive a
black hole, the higher the ratio of its radio flux to its X-ray flux.
It was thus shown in Maccarone (2004) that the best means to search
for intermediate mass black holes in globular clusters is radio,
rather than X-ray emission, and the upper limits on the mass of the
central black hole in M~15 are a factor of about 20 better with a 6.5
hour observation by the VLA than with a 3 hour observation with the
{\it Chandra} X-ray Observatory.  We note further that instrumentation
which should produce an order of magnitude improvement in radio
sensitivity is coming in the next few years (with LOFAR and the EVLA),
but no improvements to X-ray sensitivity are likely for decades (since
the next generation X-ray telescopes, Constellation-X and XEUS are not
likely to have the angular resolution to resolve out the faintest
sources in the cores of globular clusters).

Blind searches for faint (i.e. isolated or quiescent) black holes will
benefit from searches in the radio, as well.  Radio searches will have
better sensitivity than X-ray searches, and perhaps more importantly,
upcoming radio instrumentation like LOFAR (the Low Frequency Array,
see e.g. R\"ottgering 2003 or http://www.lofar.org for a description)
will have a dramatically wider field of view than current or planned
high sensitivity X-ray instrumentation. Agol \& Kamionkowski (2002)
noted that despite the fact that the brightest accreting isolated
black holes should be found in molecular clouds, where the
interstellar medium is densest, X-ray searches for such sources would
be rather inefficient since it would take too many Chandra pointings
to cover a single nearby $\sim 10$ degree large Giant Molecular Cloud.
These angular scales are well-matched, though to the field of view of
LOFAR, so deep LOFAR observations would be a very effective means of
searching for isolated black holes.

\section{Accretion mechanism and efficiency}
AK02 assumed that accretion would procede by the Bondi-Hoyle mechanism
(Bondi \& Hoyle 1944), so that\\
$\dot{M}=\lambda4\pi$$G^2M^2\rho/(v^2+c_s^2)^{3/2}$, where $\dot{M}$
is the mass accretion rate, $\lambda$ is a dimensionless parameter of
order unity, $G$ is the gravitational constanct, $M$ is the mass of
the accretor, $\rho$ is the gas density, $v$ is the relative velocity
of the accretor and the gas, and $c_s$ is the sound speed of the gas.
More recent work from a variety of sources -- numerical simulations,
observations of faint AGN and the lack of detection of large numbers
of isolated neutron stars -- has been compiled by Perna et al. (2003),
and suggests that the dimensionless constant $\lambda$ is about
$10^{-2}$ to $10^{-3}$.  A more comprehensive study of active galactic
nuclei only found that $\lambda$ varies from AGN to AGN, over the
range of 0.1 to 0.001, with 0.01 a typical value (Pellegrini 2005).
For the purposes of this paper we will make the most conservative
estimate, that $\lambda=0.001$, in order not to overpredict the number
of faint sources we expect to see.  Unless otherwise noted, we will
assume for the purposes of this paper that the black holes' space
velocities are small enough that the Bondi rate is dominated by the
temperature of the accreted gas.

Next one must estimate the efficiency at turning accreted gas into
X-rays, radio emission, and kinetic power.  AK02 simply chose an
accretion efficiency $\epsilon=10^{-5}$, where the luminosity,
$L_X=\epsilon\dot{M}c^2$.  On the other hand, given the results of
Fender, Gallo \& Jonker (2003) described above, it seems that a more
appropriate radiative efficiency is
$0.1\left(\frac{\dot{m}}{0.02}\right)$, where $\dot{m}$ is the
accretion rate in Eddington units assuming $\epsilon=0.1$.  This
corresponds to the case of equal power coming out in X-rays and jet
kinetic power at the transition between hard and soft spectral states,
which is intermediate between the most conservative (Fender et
al. 2003) and most extreme (Malzac, Merloni \& Fabian 2004) estimates
to date.

\section{Flux and source number predictions}
\subsection{Stellar mass isolated accreting black holes}
We will consider the typical appearance of stellar mass isolated
black holes accreting from the different phases of the interstellar
medium.  Obviously, the accretion rate will be the highest in the
coolest, densest phases of the ISM -- molecular gas, and the cold
neutral medium, and will be the lowest in the hottest, most tenuous
regions.  While most of the volume of the Milky Way is in the halo
where the ISM is hot and sparse, black holes should have a disk-like
distribution, albeit with a rather large scale height of about 1 kpc
(Jonker \& Nelemans 2004).  This scale height is larger than what was
found in previous studies (e.g. van Paradijs \& White 1996), primarily
because Jonker \& Nelemans (2004) found that the distances, and hence
the heights above the Galactic plane, for many black hole X-ray
binaries had been systematically underestimated.  AK02 argued that the
best assumption about the spatial distribution of the isolated black
holes would be that they follow the same distribution as the black
holes in binaries, but we note that this is likely to overestimate the
number of detectable black holes, because some of the black holes in
X-ray binaries show evidence of velocity kicks (e.g. Orosz \& Bailyn
1997; Mirabel et al. 2001).  Furthermore, the strongest velocity kicks
are most likely to unbind a binary system, and would hence give the
isolated black holes a larger scale height than the black holes in
binaries.  While a detailed population synthesis for isolated black
holes is beyond the scope of this paper, we make this point as a note
to others who might attempt work similar to AK02 in the future.

The predicted radio emission falls off less sharply with decreasing
mass accretion rate than the predicted X-ray emission.  As a result, a
very nearby, very faint isolated black hole should be more easily
detected in the radio than in the X-rays.  We also note that the
effects of interstellar extinction on the detectability of a source in
the X-rays were underestimated by AK02.  While their statement that
only about one-third of the flux from a source within a molecular
cloud will be absorbed within the 1-10 keV {\it Chandra} band, given a
typical column density of $10^{22}$ is reasonable, what is actually
important is the number of photons detected.  About 60-80\% of the
photons will be absorbed (with the lower end of the range for ACIS-I
which has a relatively hard response, and the higher end of the range
for ACIS-S), meaning that exposure times an order of magnitude longer
will be needed to reach the same sensitivity in a molecular cloud as
is needed for a low absorption column line of sight.

We summarize our assumptions: (1) $\dot{M} = 0.001 \dot{M}_{Bondi}$,
(2) $\epsilon=0.1\left(\frac{\dot{m}}{0.02}\right)$,
(3) the radio/X-ray/mass relation shown in Maccarone (2004), derived
based on the results of Merloni, Heinz \& DiMatteo (2003), (4) a flat
radio spectrum (i.e. such that the flux density $f_\nu\propto \nu^0$),
(5) black hole spatial velocities small enough such that the
temperature of the gas, rather than the velocity of the black hole
relative to the ISM is dominates the $v^2+c_s^2$ term in the Bondi
formula (6) an ISM temperature of $10^4$ K everywhere except in the
hot ISM, since for black holes greater than about 7 solar masses, the
Stromgren sphere of the X-ray source will be larger than the Bondi
radius of the black hole (7) black hole mass of $10 M_\odot$.  We then
compute the distances to which sources should be detectable with a
radio survey with a minimum flux level of 30 $\mu$Jy, which should be
achievable at 200 MHz with a 10 hour LOFAR integration.  In the dense
parts of molecular clouds (i.e. with a $n_H=10^5$), black holes should
be observable in the radio out past 10 kpc, while in the sparsest
parts of giant molecular clouds or the denser parts of the cold
neutral ISM (i.e. with a $n_H=10^3$), the black holes should still be
observable out to about 700 pc.  In the warm neutral and warm ionized
phases of the ISM (i.e. with a $n_H=0.4$), stellar mass black holes
would be observable only out to a few parsecs, and these sources
should be undetectable in the hot ionized medium - they should be
several orders of magnitude fainter than the sources in the warm
interstellar medium; any such source which would be detectable would
be at distances of order the size of the solar system, where the ISM
is not in the hot ionized phase and where, in any event,the
gravitational effects of such a system, as well as the effects of the
ionizing radiation of its progenitor would likely already be well
known.  Thus taking a 20 kpc Galactic disk with a 1 kpc height for the
$10^8$ black holes, and making the approximation that the black holes
are uniformly distributed within this volume, we find a density of
about $10^5$ BH/kpc$^3$.  It is thus unlikely that even an all-sky
radio survey will detect any isolated black holes in the warm or hot
phases of the ISM.  On the other hand, since about 5\% of the volume
of the Galactic disk is in molecular or cold neutral form, and about
$3\times10^{-4}$ of the disk is within 700 pc, about
$1.5\times10^{-5}N_{BH}$ would be detected in an all-sky survey
capable of detecting 30 $\mu$Jy sources, where $N_{BH}$ is the number
of black holes in the Galaxy.  Since LOFAR will be able to see only
about 1/4 of the sky, we can expect $\sim400$ isolated black holes to
be detectable by LOFAR.  On the other hand, the FIRST and NVSS
surveys, which go down only to $\sim1-3$ mJy levels, should be able to
detect only about 1/1000 as many sources as the LOFAR survey, and
hence it is not surprising that no isolated black holes have been
claimed from FIRST.  The LOFAR predictions are presented in Table 1.

\subsection{Isolated intermediate mass black holes}
Several different scenarios have been proposed for the distribution of
intermediate mass black holes in the galaxy.  Schneider et al. (2002)
discussed the possibility that a large fraction of the baryonic dark
matter in the Universe might be $\sim200$ solar mass black holes
formed as the remnants of Population III stars (see Bromm, Coppi \&
Larson 1999 for a discussion of the masses of Population III stars and
Fryer, Woosley \& Heger 2001 for a discussion of the resulting black
hole masses), and suggested that some of the non-nuclear bright X-ray
sources seen in other galaxies might be these black holes accreting
from the ISM.  Maccarone, Kundu \& Zepf (2003) pointed out that even
given Bondi-Hoyle accretion rates and radiatively efficient accretion,
such sources would be observable in other galaxies only if they were
in molecular clouds in contrast to some claims that the brightest
X-ray sources, even in some elliptical galaxies, might be these IMBHs.
Recently, Zhao \& Silk (2005) proposed that there should be
$\sim10^3-10^4$ mini-halos in the Milky Way, each with central black
holes of $\sim10^3-10^4 M_\odot$.  They found that these systems would
be rather difficult to detect on the basis of their X-ray emission,
but did not consider their detectability in the radio.  

The cosmological mechanisms for forming intermediate mass black holes
typically result in the black holes' being distributed in a halo,
which reduces by a factor of $\sim10$ the chance that they will be
within the disk, and also increases their velocity dispersion by a
factor of $\sim$10, hence reducing their typical luminosities by a
factor of $\sim$1000 even when they are within the disk.  Some means
have been suggested to create intermediate mass black holes in a disk
population; for example, the simulations of Portegies Zwart \&
McMillan (2002) and of G\"urkan et al. (2004) show that intermediate
mass black holes can result from runaway collisions of massive stars
in dense star clusters, which create a supermassive star which in turn
collapses into an intermediate mass black hole on a timescale of a few
Myrs after the star cluster has formed.

For intermediate mass black holes, the radio band should, indeed, be a
much more efficient means of detecting sources than the X-rays are.
That intermediate mass or higher black holes accreting at low
luminosities should be more easily detected in the radio than in the
X-rays is a point already made in Maccarone (2004).  In Maccarone,
Fender \& Tzioumis (2005), it was estimated that the dependence on the
radio luminosity on the black hole mass, for a given interstellar
medium density and temperature, was roughly
$L_R\propto$$M_{BH}^{3.2}$.  A factor of 25 change in black hole mass
from stellar mass black holes to the black holes expected from
Population III star remnants should result in a factor of about 30000
change in the radio luminosity.  In the dense parts of a molecular
cloud, such a source would reach a luminosity of $10^{37}-10^{38}$
ergs/sec in the X-rays, and a radio flux density of about 150 mJy for
a distance out to 10 kpc; such sources would already be known in
nearly any of the Milky Way's molecular clouds, and would be
detectable throughout the Local Group.  Even in the less dense parts
of a molecular cloud, these sources would appear in the NVSS and FIRST
catalogs out to about 5 kpc, though they would be only at luminosities
of about $10^{34}$ ergs/sec in the X-rays.  The fact that no excess of
unusual radio sources has been reported from NVSS measurements of
molecular clouds would seem to indicate that less than about 100 IMBHs
exist in a disk population in the Milky Way.  LOFAR would be expected
to see such systems in any part of the Galaxy's molecular or neutral
ISM it observes (i.e. out to about 40 kpc), or in the warm neutral ISM
within about 500 pc.

The detection of a halo distribution of sources would be considerably
harder, primarily because the halo sources would be moving with much
higher velocities with respect to the Galactic disk. Assuming a
typical velocity of 100 km/sec, and a black hole mass of 2600
$M_\odot$ (as in e.g. Zhao \& Silk 2005), we find that these systems
should still be observable with LOFAR (i.e. above 30 $\mu$Jy) in the
radio out to about 15 kpc, if in the molecular or cold neutral part of
the ISM.  We expect these objects to be about 10 times fainter in
radio than slower, smaller black holes in similar environments at
similar distances.  On the other hand, we expect them to be about
$10^3$ times fainter in the X-rays than the similar smaller, slower
black holes, and hence to be undetectable in X-rays except in the
nearest clouds and even then only with rather long {\it Chandra}
observations.  With FIRST, such sources would be seen out to only
about 1 kpc, in molecular clouds, so the current lack of detections is
not surprising.  With LOFAR, the expect number of sources seen would
be about $\frac{1}{4}\times3000\times\frac{1}{100}$, or about 8, since
about 1/4 of the Galaxy is observable, and about 1\% of the volume of
the galaxy is in the cold dense phase and Zhao \& Silk (2005)
estimated about 3000 IMBHs.  This prediction is presented as the last
entry in Table 1.

\subsection{Quiescent accreting black hole systems}
Calculations of the expected number of black hole low mass X-ray
binaries are quite sensitive to many as yet not well known parameters,
including, but not limited to the velocity kick distribution of black
holes, the initial stellar mass function, the binary fraction, the
initial distribution function of binary periods, the distribution
function of mass ratios of stars in binaries, the details of common
envelope evolution and the details of binary stellar evolution (see
e.g. the population synthesis codes described in Portegies Zwart \&
Verbunt 1996; Belczynski, Kalogera \& Bulik 2002; Pfahl, Rappaport \&
Podsiadlowski 2003).  Given all these unknowns, along with the
possibility that the physics in the codes might be incomplete, any
additional observational constraints on the number of these sources
would be useful.  Typical numbers of low mass black hole binaries
predicted are a $\sim3000$ (see e.g. Portegies Zwart, Verbunt \&
Ergma 1997), to a $\sim30000$ (Romani 1992), which would mean that the
kinetic energy input from quiescent systems may rival that of
outbursting systems.  Additional low luminosity systems might exists
in the form of faint wind-fed black holes (Belczynski \& Taam 2004),
but the number of such sources are is quite uncertain.

At the present time, there are no black holes in binaries that are
known to exist except those which have been detected through their
X-ray emission.  The last serious attempt to find such objects was an
attempt to measure the masses of binary companions in normal binary
star systems that appeared to have massive counterparts, and failed to
establish the presence of compact objects in these systems (Trimble \&
Thorne 1969).  The typical luminosity of a quiescent black hole X-ray
binary in the X-rays is $10^{32}$ ergs/sec, based on making a factor
of 5 bolometric correction (as established in Portegies Zwart, Dewi \&
Maccarone 2004) from the median $2\times10^{31}$ ergs/sec found in the
Chandra bandpass (Tomsick et al. 2003).  The ROSAT All-sky survey was
capable of detecting {\it unabsorbed} quiescent black holes out to
about 1 kpc, but we hasten to note that ROSAT's sensitivity falls off
sharply in the Galactic plane, where these objects are typically
found.

Several more of these sources might be detected in the radio than have
already been detected in X-rays, as well.  Quiescent black hole
binaries should be detectable at $5\sigma$ in one hour of LOFAR time
at 200 MHz out to a distance of 500 pc.  It should be possible, too,
though, for LOFAR to make much deeper all-sky surveys than an hour on
each field of view.  Assuming the same 30 $\mu$Jy limit as for the
other types of sources, LOFAR should detect most of the quiescent
black hole binaries in the Galaxy.  Futhermore, LOFAR will have better
angular resolution than ROSAT, allowing many more of the faint sources
to be associated with a unique optical counterpart.  Finally,
quiescent X-ray binaries can show factor of $\sim10$ variability in
the radio (e.g. Hjellming et al. 2000), which would allow one to
follow up only a fraction of the LOFAR sources.  The ROSAT all-sky
survey does not provide the kind of variability data that would be
needed to make the follow-up more feasible.

\begin{table*}
\begin{tabular}{lllllllll}
\hline
Phase/type &$M_{BH}$&$n_H$&$T_{ISM}$&$N_{BH}$&$L_X$&$d_{radio}$&$N_{radio}$\\
\hline
GMC Core& 10&$10^5$&$10^4$&$\sim1$ &5$\times10^{33}$&12&$\sim$1\\
GMC/cold neutral&10&$10^3$&$10^4$&1.3$\times10^6$&5$\times10^{29}$&0.7&400\\
warm ISM&10&0.4&$10^4$&$5\times10^7$&7$\times10^{22}$&.005&0\\
hot ISM&10&0.01&$10^6$&$5\times10^7$&5$\times10^{13}$& $10^{-5}$&0\\
GMC/cold, fast halo IMBH& 2600&$10^3$&$10^4$&30&$8\times10^{30}$&15&10\\ 
IMBH/disk pop/cold ISM&260&$10^3$&$10^4$&*&$8\times10^{33}$&40&*\\
IMBH/disk pop/GMC&260&$10^5$&$10^4$&*&$8\times10^{37}$&800&*\\
IMBH/disk pop/warm ISM&260&$0.4$&$10^4$&*&$1\times10^{27}$&0.5&*\\
\hline
\end{tabular}
\caption{The calculated results for the types of black holes discussed
in the text, assuming $10^8$ total isolated black holes in the Galaxy
and that 1/4 of the sky will be seen by LOFAR.  The columns are: phase
of the ISM and black hole type, mass of the black hole in $M_\odot$,
$n_H$ in cm$^{-3}$, temperature of the ISM in Kelvins, the number of
black holes expected to be in that phase of the ISM, the predicted
X-ray luminosity in ergs/sec, the distance in kpc at which the radio
flux would be greater than 30 $\mu$Jy, and the expected number of
sources in that phase of the ISM within $d_{radio}$.  The asterisks
are in place of the numbers of existing and observable intermediate
mass black holes with disk-like populations because there exists no
prediction for how many of these objects should exist.  Only the
objects we expect not to detect are predicted to be fainter than the
range where the Merloni et al. (2003) correlation has been shown to
work.}
\end{table*}

\section{Caveats, comparisons with previous work and conclusions}

A few words of caution are warranted here.  The velocities of the
black holes may be a bit higher than we have estimated them to be, and
as the Bondi accretion rate scales with the velocity of the black hole
cubed, this can have a profound impact on the estimated fluxes in any
waveband.  On the other hand, we have made a conservative assumption
about the fraction of the Bondi rate that is accreted, which should
mitigate the effects of possibly underestimating the space velocities.
If the typical value of $\lambda$ is 0.01, as suggested by Pellegrini
(2005), then we have underestimated the typical X-ray luminosities by
a factor of about 100, and the typical radio luminosities by a factor
of 25.  This would lead to an increase in the expected number of
sources by a factor of 25 (if we are already seeing sources out to the
full scale height of the Galaxy) to 125 (if the maximum distance for
sources to be observable is much less than the scale height of the
Galaxy).

Regarding the effects of black hole mass, the relationship of Merloni
et al. (2003) has considerable scatter for causes which are not clear;
some suggestions have been measurement errors, beaming effects (Heinz
\& Merloni 2004), and effects of the spin of the black hole and
spectral state of the accretion disk (Maccarone, Gallo \& Fender
2003).  On the other hand, for the stellar mass black holes, the
relationship of Gallo et al. (2003) fits a rather tight correlation,
so we can be reasonably confident that we have handled the stellar
mass black hole sources correctly.  Mass spread among stellar black
holes is likely to increase the number of detectable sources, since
increasing the spread in flux without changing the mean will increase
the number in the high flux tail.

Most previous papers about the search for accreting isolated black
holes have focused on how to detect these systems in the X-rays.  Here
we have focused primarily on how one might detect these systems in the
radio, rather than in the X-rays for two reasons.  The first is that
previous studies have generally systematically overestimated the flux
levels one would expect in the X-rays from isolated accreting black
holes.  AK02 made the most conservative predictions to date, assuming
the full Bondi rate, but with a radiative efficiency of $10^{-5}c^2$,
rather than the standard 0.1$c^2$ used for radiatively efficient
accretion.  However, since the publication of AK02, it has become
clear both that the Bondi rate overestimates the true accretion rate
from the interstellar medium by 1-3 orders of magnitude (Perna et
al. 2003; Pellegrini 2005).  It has also become clear that the
radiative efficiency factor drops with decreasing accretion rate,
although this has been suggested for quite some time now (see
e.g. Ichimaru 1977; Narayan \& Yi 1994).  As a result, our predicted
X-ray luminosities from these systems are orders of magnitude lower
than those predicted by AK02.

Radio surveys should be superior to X-ray surveys in searching for
IMBHs for another reason.  AK02 noted that while molecular clouds are
clearly the locations where isolated black holes will be easiest to
detect, surveying them would be nearly impossible due to the
relatively small fields of view of existing sensitive X-ray telescopes
like {\it Chandra} and {\it XMM}; to cover the typical 10 degree field
of view of a nearby giant molecular cloud would require about 1000
pointings with {\it Chandra} or {\it XMM}, and to reach reasonable
sensitivity limits would require at least 10 kiloseconds - that is to
say that it would take about a month to cover a single molecular cloud
effectively.  Furthermore, they underestimated the effects of
interstellar absorption on the sensitivity of these instruments
because they computed the fraction of the energy flux absorbed, rather
than the fraction of the photons absorbed, and the soft X-rays are, of
course, preferentially absorbed.  In fact, it would take about a year
of Chandra time to do a good survey for isolated black holes in a
single giant molecular cloud.  Because of LOFAR's large field of view,
a year of LOFAR time at 200 MHz is sufficient to observe the whole
Northern sky for about 300 hours per field, which formally should be
enough to reach a noise level of 2$\mu$Jy, but which might approach to
the confusion limit or other systematic limits.

We have shown that upcoming deep, wide-field radio surveys will be
substantially more effective for finding isolated black holes than any
other means which has been proposed to date.  We have also discussed
the kinetic and cosmic ray luminosities likely to come from these
sources and other faint accreting black holes and how the Galactic
cosmic ray and kinetic luminosities can be used to constrain the
numbers of such sources.

\section{acknowledgments}
I am grateful for useful discussions with Chris Belczynski, Rob
Fender, Ed van den Heuvel, Christian Kaiser, Simon Portegies Zwart,
Michael Sipior, Ben Stappers, Ron Taam, Wing-Fai Thi, Ralph Wijers,
Rudy Wijnands.  I thank the referee Andrea Merloni for useful
suggestions.

\label{lastpage} 

\end{document}